\documentclass{PoS}

\title{Progress on Calorons}

\newcommand{\beqa}{\begin{eqnarray}}
\newcommand{\eeqa}{\end{eqnarray}}
\newcommand{\bea}{\begin{array}}
\newcommand{\eea}{\end{array}}
\newcommand{\Pexp}{{\rm Pexp}}
\newcommand{\cA}{{\cal{A}}}
\newcommand{\cP}{{\cal{P}}}
\newcommand{\Tr}{{\rm Tr}}
\newcommand{\tr}{{\rm tr}}
\newcommand{\diag}{{\rm diag}}
\newcommand{\pl}{{{\cal P}_\infty}}
\newcommand{\half}{{\scriptstyle{\frac{1}{2}}}}
\newcommand{\inv}{{-1}}

\ShortTitle{Progress on Calorons}

\author{\speaker{Pierre van Baal}\\
        Instituut-Lorentz for Theoretical Physics\\
        University of Leiden, P.O.Box 9506\\
        NL-2300 RA Leiden, The Netherlands\\
        E-mail: \email{vanbaal@lorentz.leidenuniv.nl}}

\abstract{The progress on calorons (finite temperature instantons)
is sketched. In particular there is some interest for confining
temperatures, where the holonomy is non-trivial.}

\FullConference{8th Conference Quark Confinement and the Hadron Spectrum \\
		 September 1-6 2008\\
		 Mainz, Germany}

\begin{document}

\section{Introduction}

There has been a revised interest in studying instantons at finite temperature 
$T$, so-called calorons~\cite{HaSh,GPY}, because new explicit solutions could 
be obtained where the Polyakov loop at spatial infinity (the so-called 
holonomy) is non-trivial. They reveal more clearly the monopole constituent 
nature of these calorons~\cite{PLB}. Non-trivial holonomy is therefore 
expected to play a role in the confined phase (i.e. for $T<T_c$) where the 
trace of the Polyakov loop fluctuates around small values. The properties of 
instantons are therefore directly coupled to the order parameter for the 
deconfining phase transition. 

At finite temperature $A_0$ plays in some sense the role of a Higgs field 
in the adjoint representation, which explains why magnetic monopoles occur 
as constituents of calorons. Since $A_0$ is not necessarily static it is 
better to consider the Polyakov loop as the analog of the Higgs field,
$P(t,\vec x)=\Pexp\left(\int_0^\beta A_0(t+s,\vec x)ds\right)$,
which transforms under a periodic gauge transformation $g(x)$ to $g(x)P(x) 
g^\inv(x)$, like an adjoint Higgs field. Here $\beta=1/kT$ is the period in 
the imaginary time direction, under which the gauge field is assumed to be 
periodic. Finite action requires the Polyakov loop at spatial infinity to be
constant. For SU($n$) gauge theory this gives $\pl=\lim_{|\vec x|\to\infty} 
P(0,\vec x)=g^\dagger\exp(2\pi i\diag(\mu_1,\mu_2,\ldots,\mu_n))g$, where 
$g$ brings $\pl$ to its diagonal form, with $n$ eigenvalues being ordered 
according to $\sum_{i=1}^n\mu_i=0$ and $\mu_1\leq\mu_2\leq\ldots\leq\mu_n
\leq\mu_{n+1}\equiv1+\mu_1$. In the algebraic gauge, where $A_0(x)$ is 
transformed to zero at spatial infinity, the gauge fields satisfy the boundary 
condition $A_\mu(t+\beta,\vec x)=\pl A_\mu(t,\vec x)\cP_\infty^\inv$.

Caloron solutions are such that the total magnetic charge vanishes. A single 
caloron with topological charge one contains $n-1$ monopoles with a unit 
magnetic charge in the $i$-th U(1) subgroup, which are compensated by the 
$n$-th monopole of so-called type $(1,1,\ldots,1)$, having a magnetic charge 
in each of these subgroups~\cite{KvBn}. At topological charge $k$ there are 
$kn$ constituents, $k$ monopoles of each of the $n$ types. Monopoles of type 
$j$ have a mass $8\pi^2\nu_j/\beta$, with $\nu_j\equiv\mu_{j+1}-\mu_j$. The 
sum rule $\sum_{j=1}^n\!\nu_j\!=\!1$ guarantees the correct action, $8\pi^2 k$. 

Prior to their explicit construction, calorons with non-trivial holonomy were 
considered irrelevant~\cite{GPY}, because the one-loop correction gives rise 
to an infinite action barrier. However, the infinity simply arises due to the 
integration over the finite energy density induced by the perturbative 
fluctuations in the background of a non-trivial Polyakov loop~\cite{Weiss}.
The calculation of the non-perturbative contribution was performed 
in~\cite{Diak}. When added to this perturbative contribution, with minima at 
center elements, these minima turn unstable for decreasing temperature right 
around the expected value of $T_c$. This lends some support to monopole 
constituents being the relevant degrees of freedom which drive the transition 
from a phase in which the center symmetry is broken at high temperatures to 
one in which the center symmetry is restored at low temperatures. Lattice 
studies, both using cooling~\cite{Berlin} and chiral fermion 
zero-modes~\cite{Gatt} as filters, have also conclusively confirmed 
that monopole constituents do dynamically occur in the confined phase. 
 
\section{Some Properties of Caloron Solutions}

Using the classical scale invariance we can always arrange $\beta=1$, as will
be assumed throughout. A remarkably simple formula for the $SU(n)$ action
density exists~\cite{KvBn},
\beqa
\Tr F_{\alpha\beta}^{\,2}(x)=\partial_\alpha^2\partial_\beta^2\log\psi(x),
\quad\psi(x)=\half\tr(\cA_n\cdots \cA_1)-\cos(2\pi t), \nonumber\\ 
\cA_m\equiv\frac{1}{r_m}\left(\!\!\!\bea{cc}r_m\!\!&|\vec\rho_{m+1}|
\\0\!\!&r_{m+1}\eea\!\!\!\right)\left(\!\!\!\bea{cc}\cosh(2\pi\nu_m r_m)\!\!&
\sinh(2\pi\nu_m r_m)\\ \sinh(2\pi\nu_m r_m)\!\!&\cosh(2\pi\nu_m r_m)\eea
\!\!\!\right),\nonumber
\eeqa
with $r_m\equiv|\vec x-\vec y_m|$ and $\vec\rho_m\equiv\vec y_m-\vec 
y_{m-1}$, where $\vec y_m$ is the location of the $m^{\rm th}$ constituent 
monopole with a mass $8\pi^2\nu_m$. Note that the index $m$ should be 
considered mod $n$, such that e.g. $r_{n+1}=r_1$ and $\vec y_{n+1}=\vec y_1$ 
(there is one exception, $\mu_{n+1}=1+\mu_1$). It is sufficient that only one 
constituent location is far separated from the others, to show that one can 
neglect the $\cos(2\pi t)$ term in $\psi(x)$, giving rise to a static action 
density in this limit~\cite{KvBn}.

\begin{figure}[htb]
\vspace{3.0cm}
\includegraphics{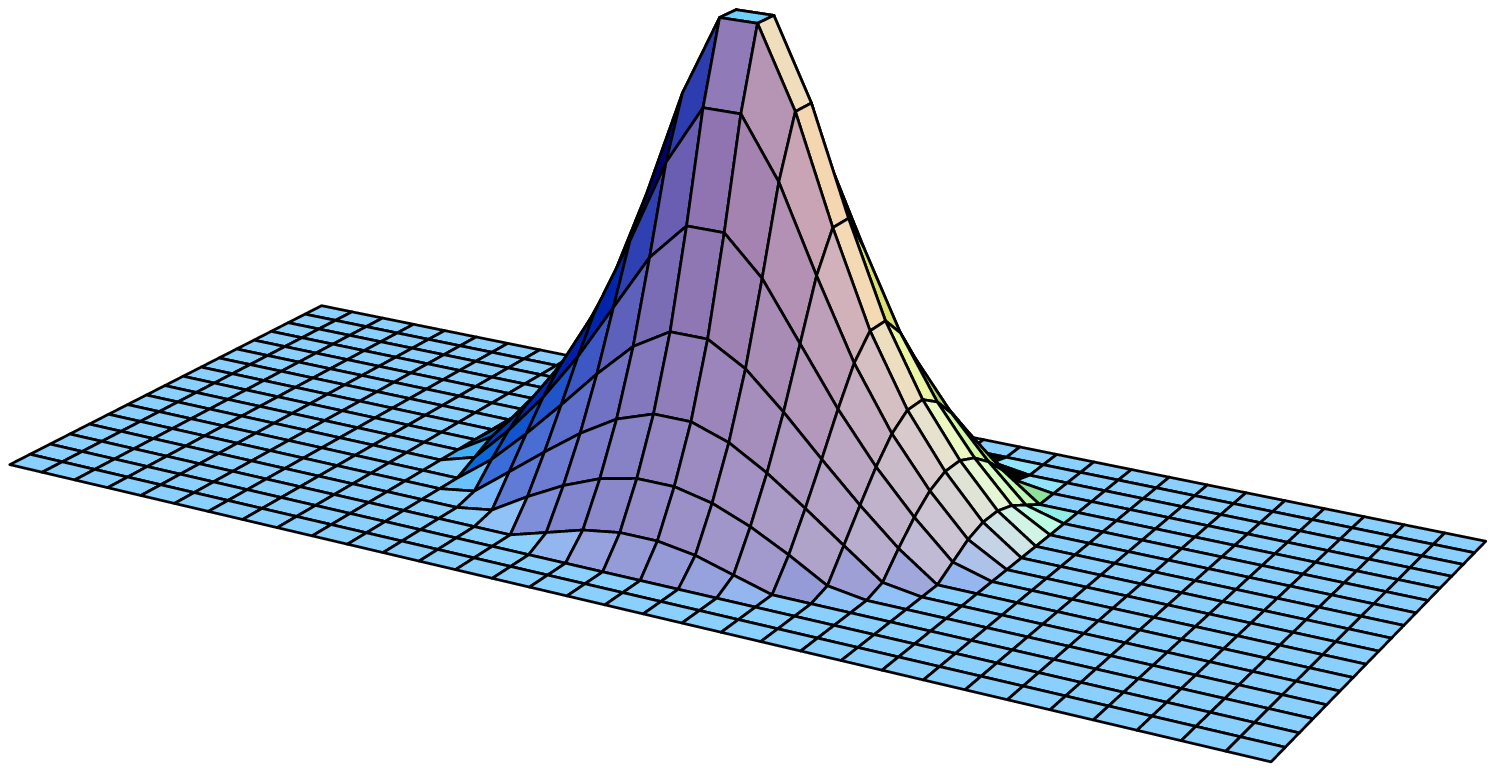}
\includegraphics{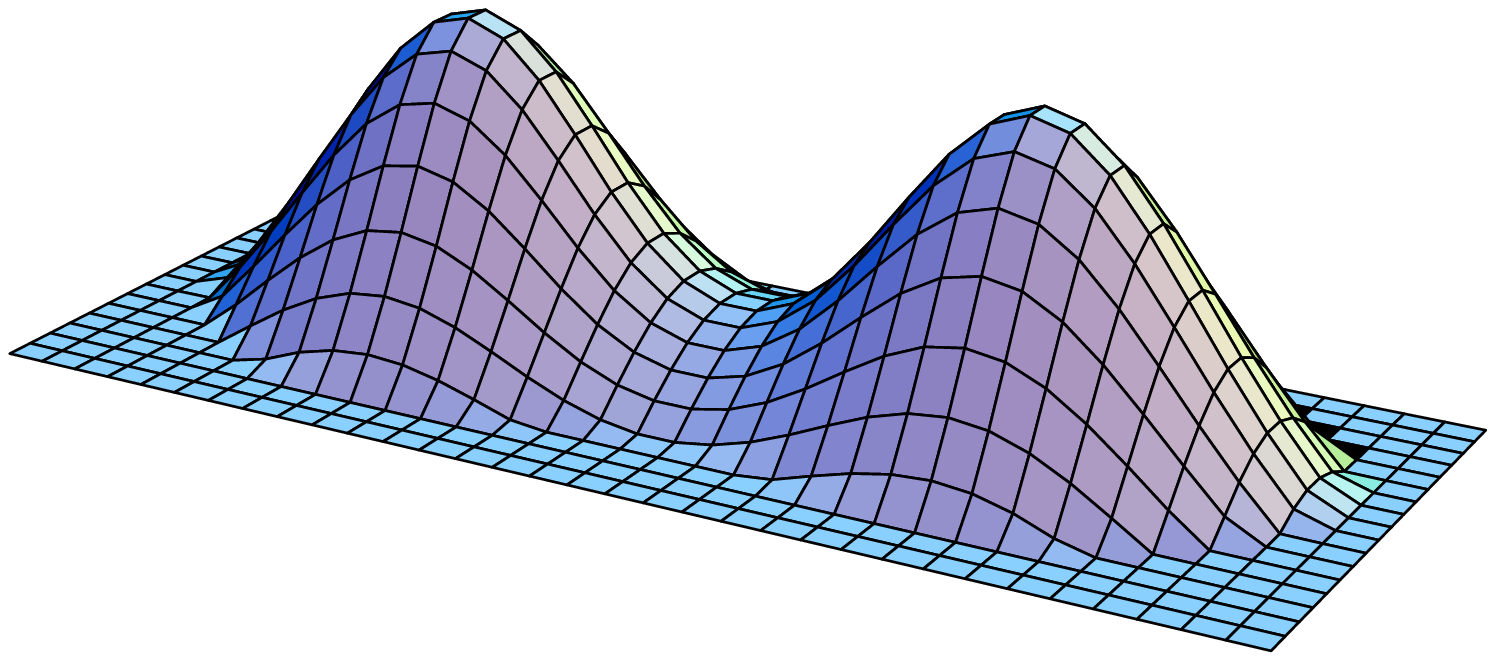}
\includegraphics{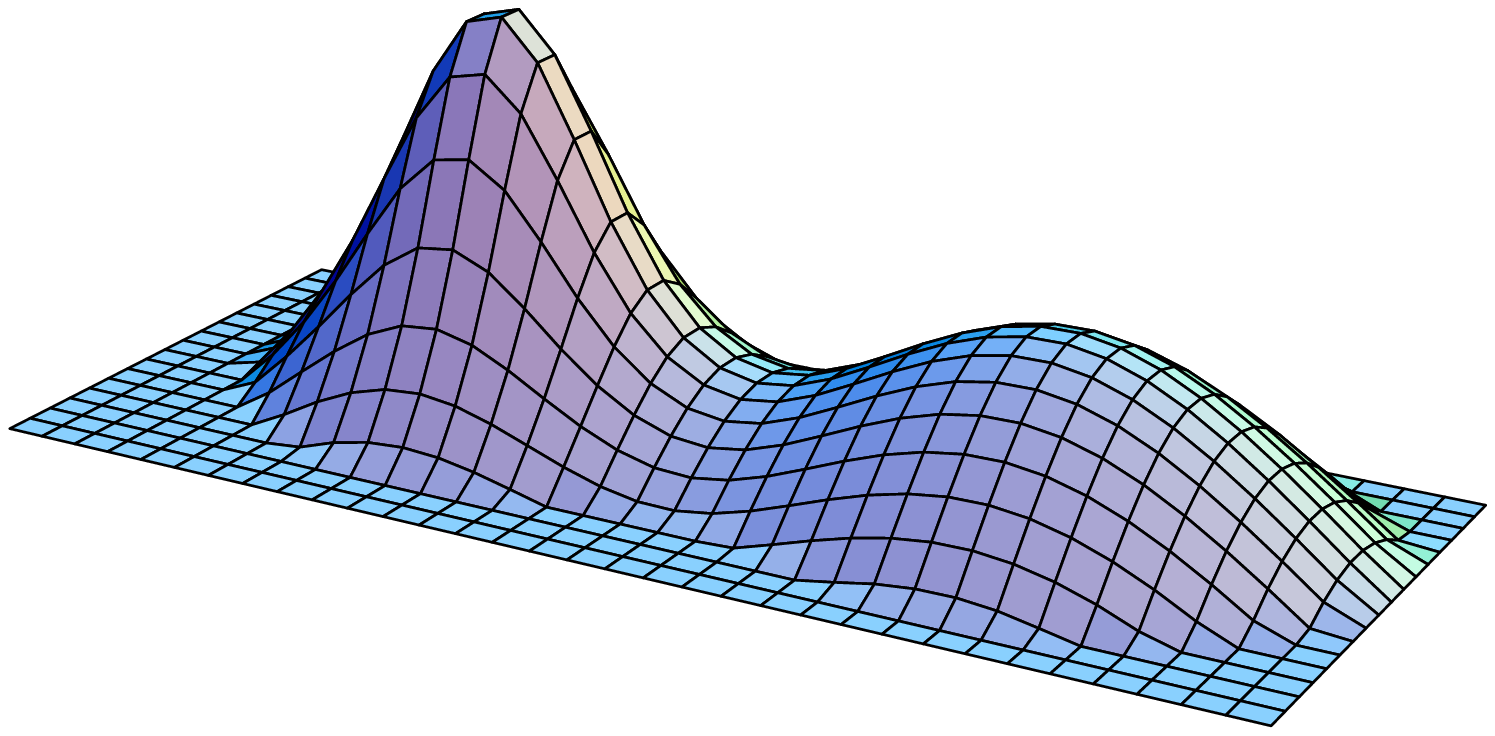}
\caption{Shown are three charge one SU(2) caloron profiles at $t=0$ with 
$\beta=1$ and $\rho=1$. From left to right for $\mu_2=-\mu_1=0$ ($\nu_1=0,
\nu_2=1$), $\mu_2=-\mu_1=0.125$ ($\nu_1=1/4,\nu_2=3/4$) and $\mu_2=-\mu_1=
0.25$ ($\nu_1=\nu_2=1/2$) on equal logarithmic scales, cutoff below an action 
density of $1/(2e)$.}
\label{fig:fig1}
\end{figure}

In Fig.~1 we show how for SU(2) there are two lumps, except that the second 
lump is absent for trivial holonomy. Fig.~2 demonstrates for SU(2) and SU(3) 
that there are indeed $n$ lumps (for SU($n$)) which can be put anywhere. These 
lumps are constituent monopoles, where one of them has a winding in the 
temporal direction (which cannot be seen from the action density).

\begin{figure}[htb]
\vspace{4.3cm}
\includegraphics{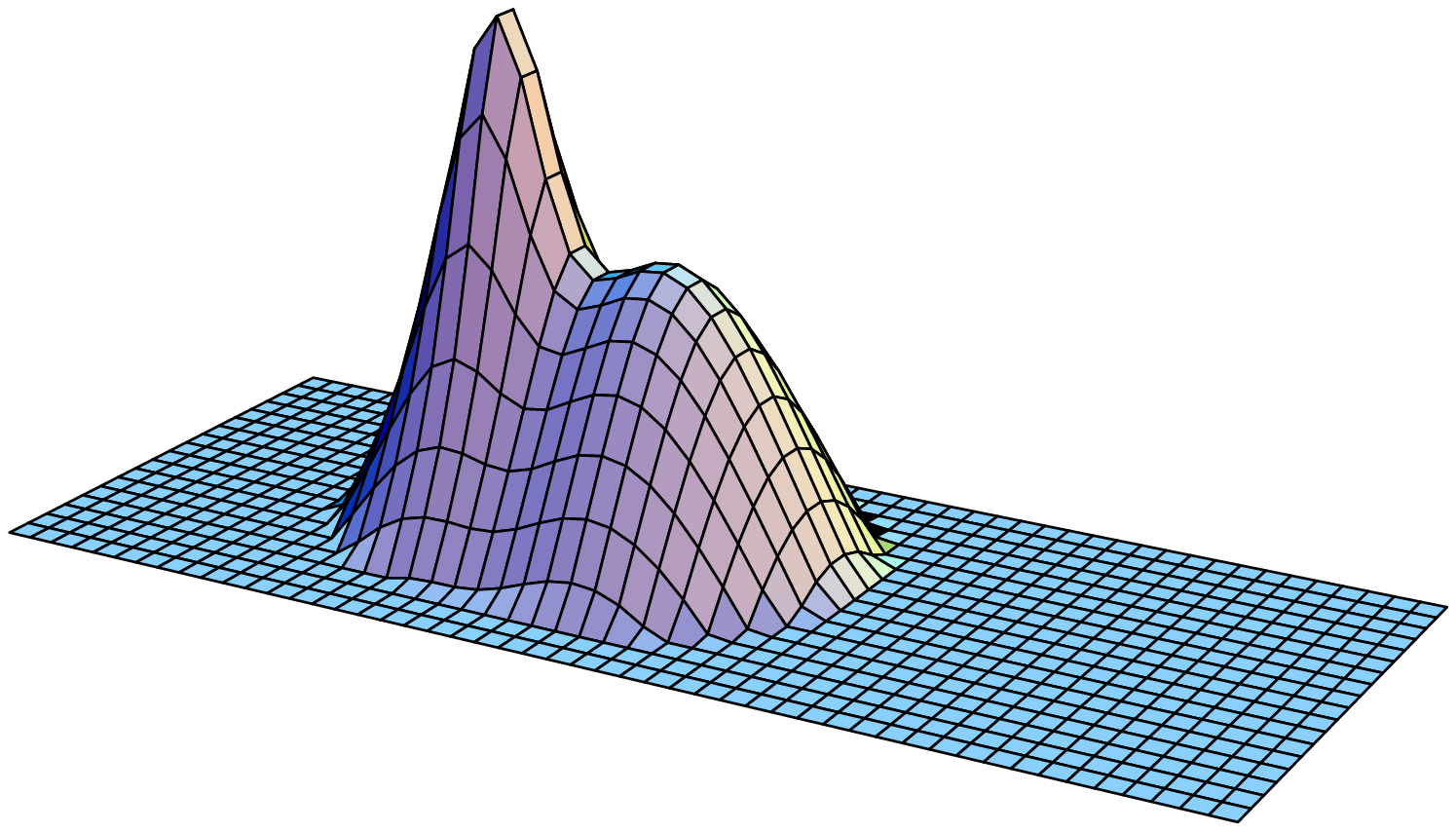}
\includegraphics{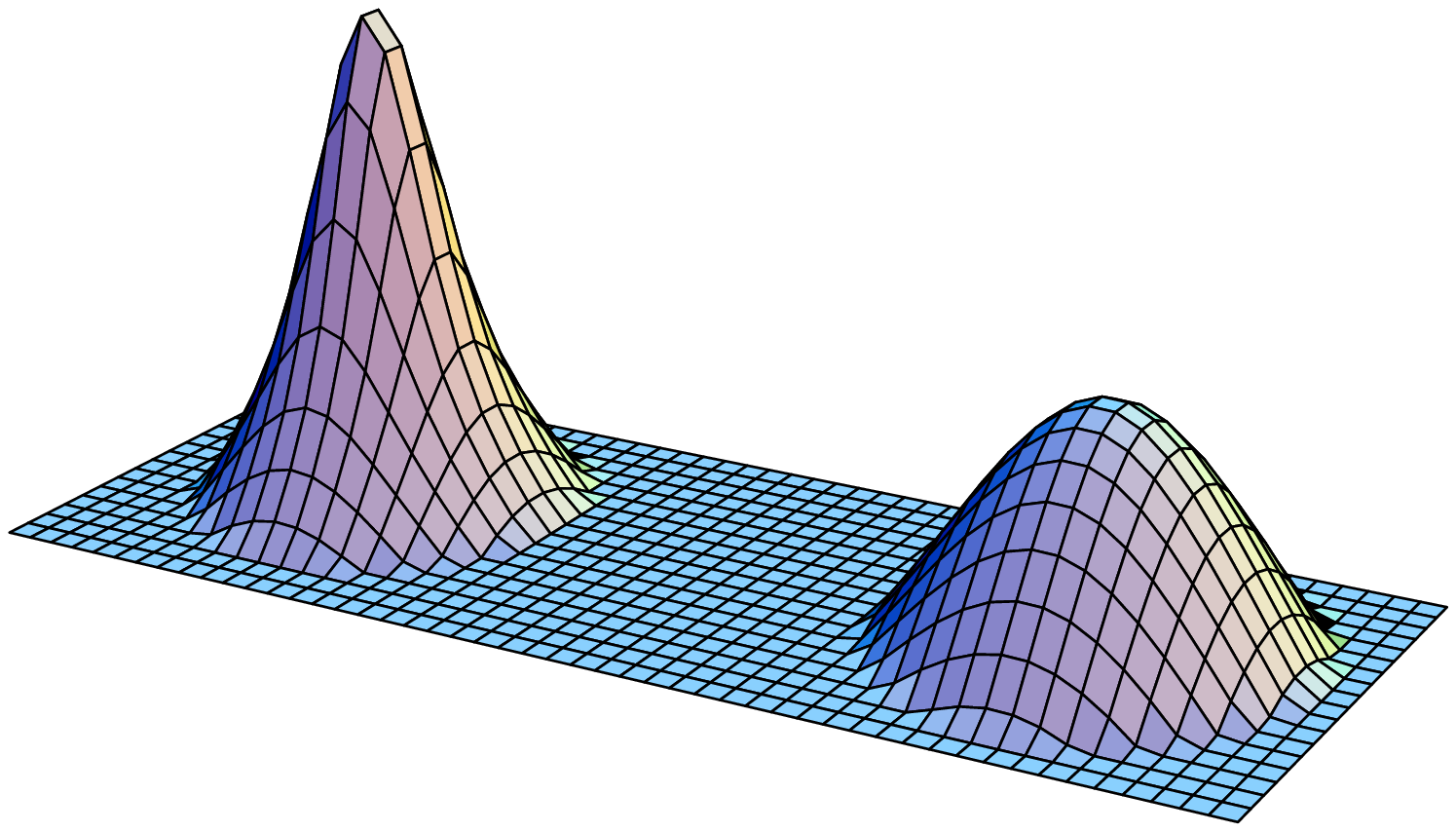}
\includegraphics{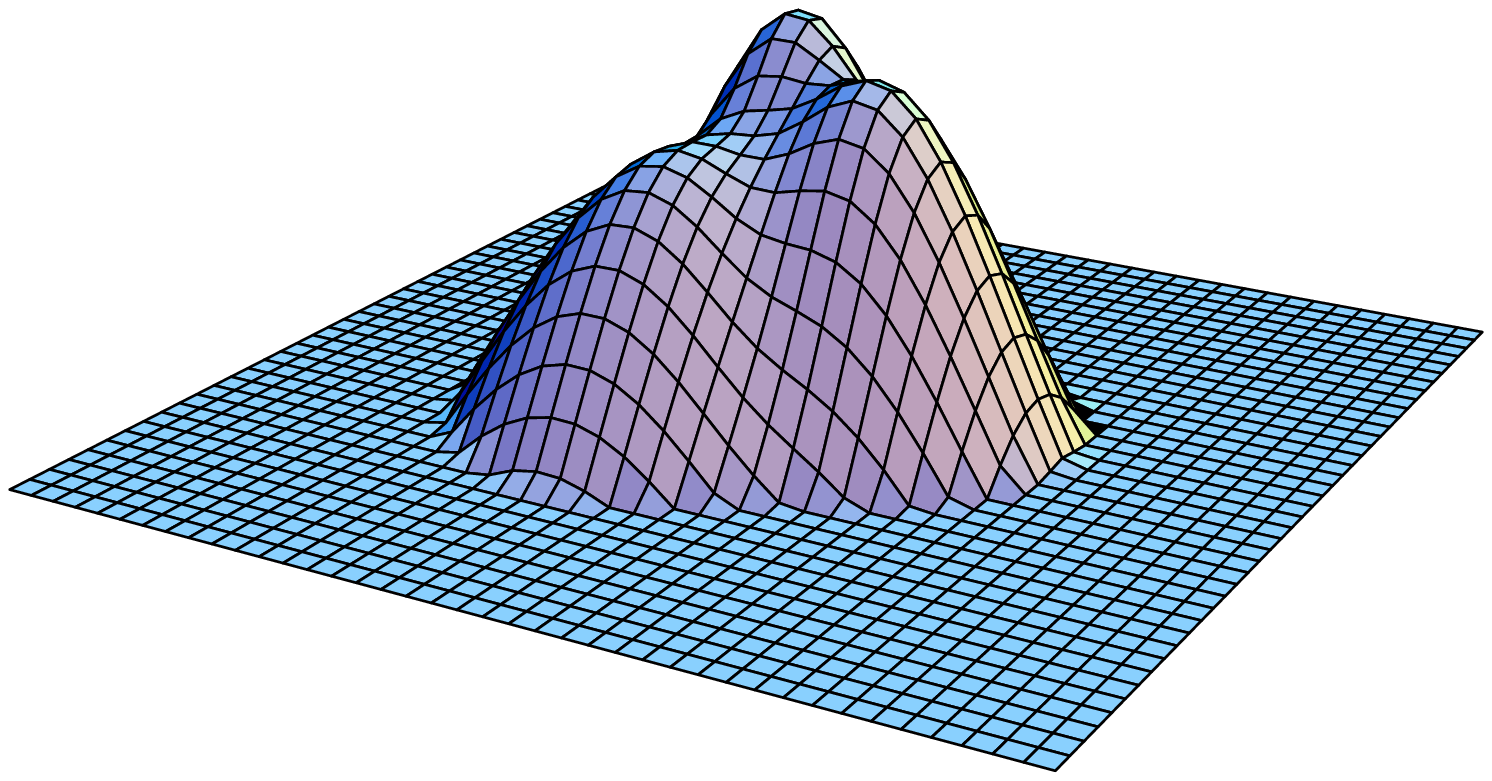}
\includegraphics{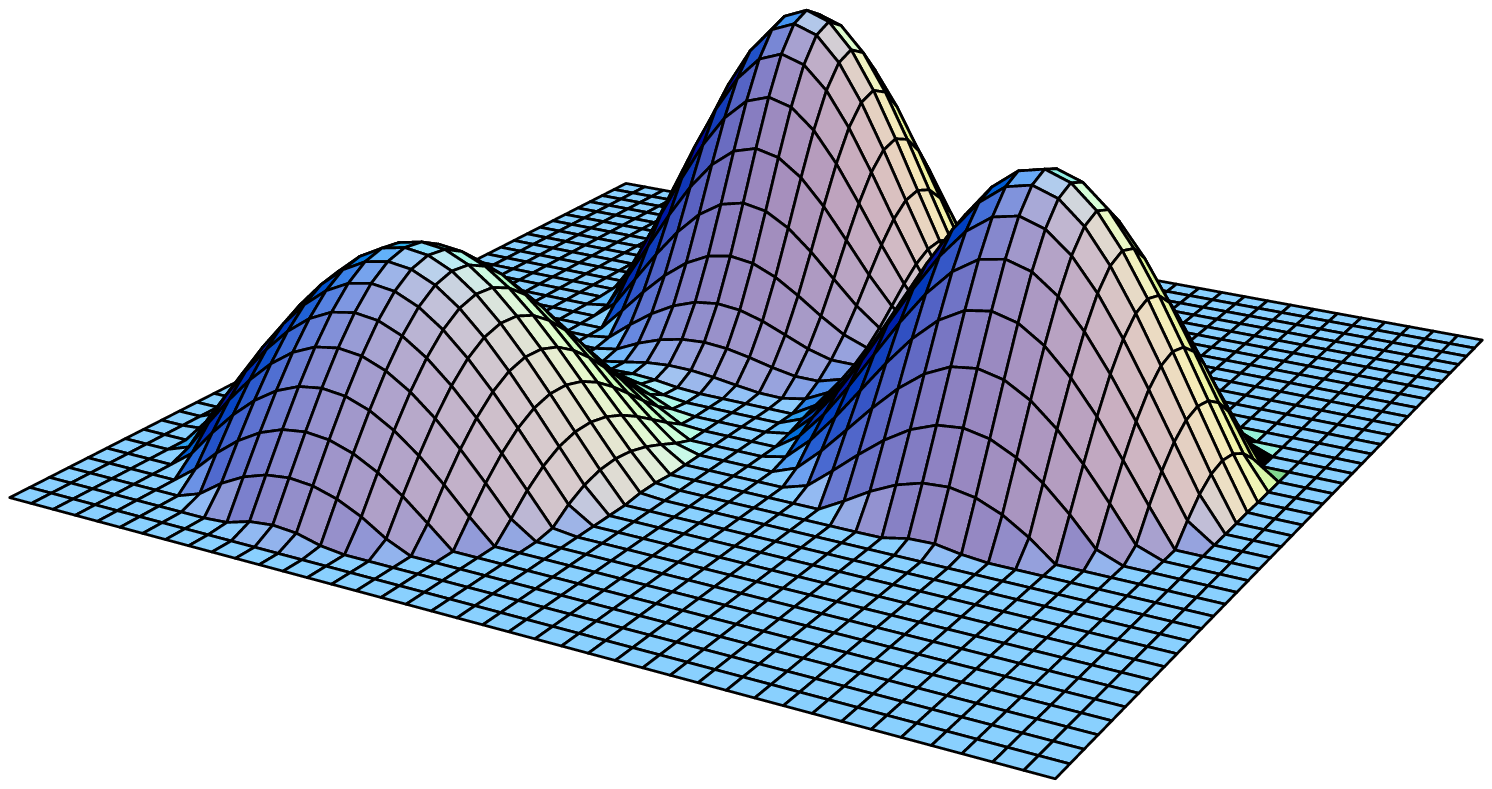}
\caption{On the left are shown two charge one SU(2) caloron profiles at $t=0$
with $\beta=1$ and $\mu_2=-\mu_1=0.125$, for $\rho=1.6$ (bottom) and 0.8 (top) 
on equal logarithmic scales, cutoff below an action density of $1/(2e^2)$. 
On the right are shown two charge one SU(3) caloron profiles at $t=0$ and 
$(\nu_1,\nu_2,\nu_3)=(1/4,7/20,2/5)$, implemented by $(\mu_1,\mu_2,\mu_3)=
(-17/60,-1/30,19/60)$. The bottom configuration has the location of the 
lumps scaled by $8/3$. They are cutoff at $1/(2e)$.}
\label{fig:fig2}
\end{figure}

\subsection{Fermion Zero-Modes}

An essential property of calorons is that the chiral fermion zero-modes are 
localized to constituents of a certain charge only. The latter depends on the 
choice of boundary condition for the fermions in the imaginary time direction 
(allowing for an arbitrary U(1) phase $\exp(2\pi iz)$)~\cite{ZM}. This provides
an important signature for the dynamical lattice studies, using chiral fermion 
zero-modes as a filter~\cite{Gatt}. To be precise, the zero-modes are localized
to the monopoles of type $m$ provided $\mu_m<z<\mu_{m+1}$. Denoting the 
zero-modes by $\hat\Psi_z(x)$, we can write $\hat\Psi^\dagger_z(x)\hat\Psi_z(x)
=-(2\pi)^{-2}\partial_\mu^2\hat f_x(z,z)$, where $\hat f_x(z,z')$ is a Green's 
function which for $z\in[\mu_m,\mu_{m+1}]$ satisfies $\hat f_z(z,z)=\pi
<\!\!v_m(z)|\cA_{m\!-\!1}\cdots \cA_1\cA_n\cdots \cA_m|w_m(z)\!\!>/(r_m\psi)$,
where the spinors $v_m$ and $w_m$ are defined by $v_m^1(z)=-w_m^2(z)=\sinh
\left(2\pi(z\!-\!\mu_m)r_m\right)$, and $v_m^2(z)=w_m^1(z)=\cosh\left(2\pi
(z\!-\!\mu_m)r_m\right)$.

To obtain the finite temperature fermion zero-mode one puts $z=\half$, whereas 
for the fermion zero-mode with periodic boundary conditions one takes $z=0$.
From this it is easily seen that in case of well separated constituents the
zero-mode is localized only at $\vec y_m$ for which $z\in[\mu_m,\mu_{m+1}]$.
To be specific, in this limit $\hat f_x(z,z)=\pi\tanh(\pi r_m\nu_m)/r_m$
for $SU(2)$, and more generally $\hat f_x(z,z)=2\pi\sinh[2\pi(z-\mu_m)r_m]
\sinh[2\pi(\mu_{m+1}-z)r_m]/(r_m\sinh[2\pi\nu_mr_m])^{-1}$. We illustrate 
in Fig.~3 the localization of the fermion zero-modes for the case of $SU(3)$. 

\begin{figure}[htb]
\vspace{3.1cm}
\includegraphics{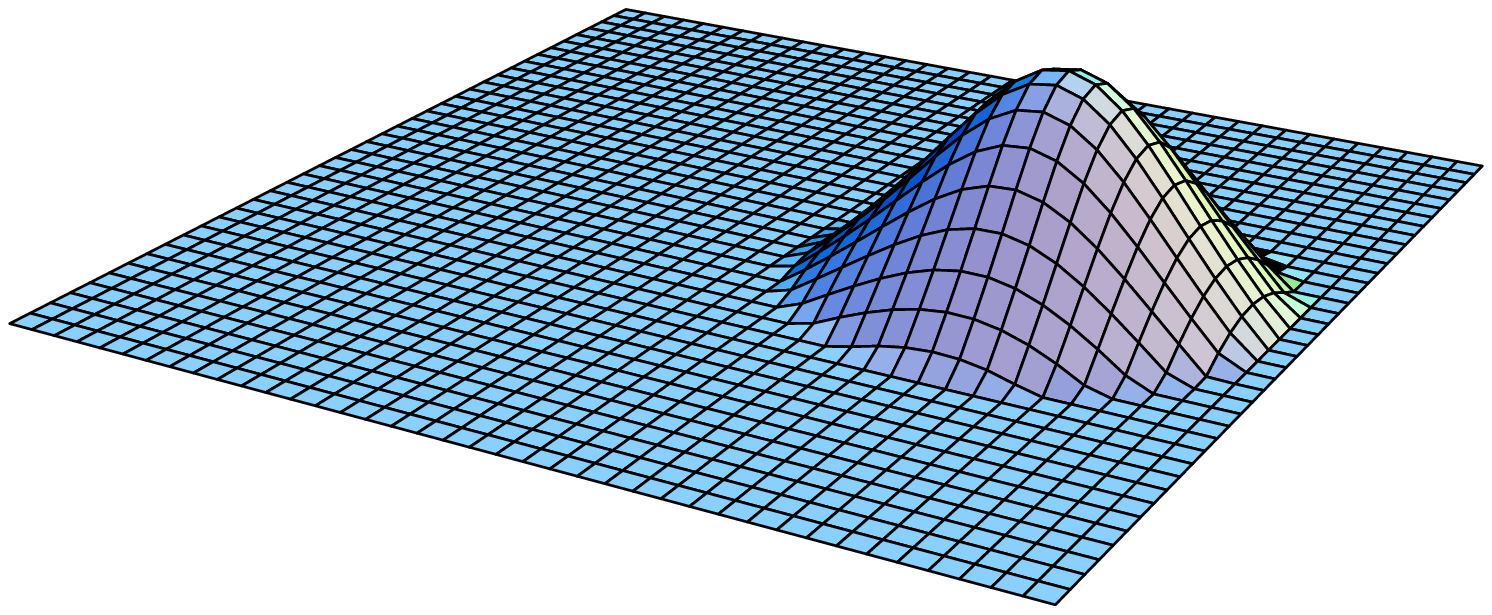}
\includegraphics{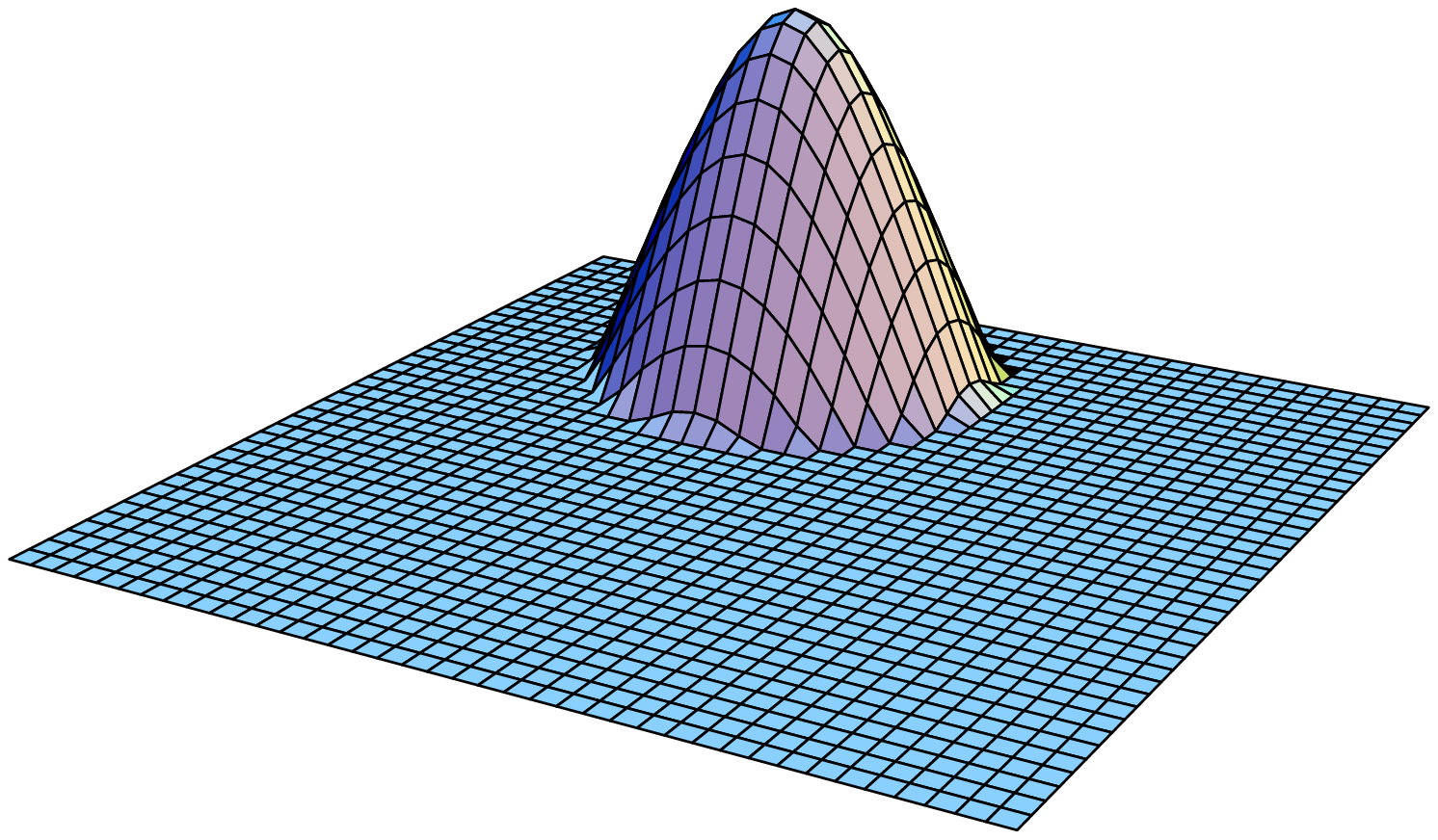}
\caption{For the SU(3) configuration in the lower right corner of Fig.~2 we 
have determined on the left the zero-mode density for fermions with 
anti-periodic boundary conditions in time and on the right for periodic 
boundary conditions. They are plotted at equal logarithmic scales, cut off
below $1/e^5$.}\label{fig:fig3}
\end{figure}

\subsection{Calorons of Higher Charge}

We have been able to use a ``mix'' of the ADHM and Nahm formalism~\cite{ADHM},
both in making powerful approximations, like in the far field limit (based on
our ability to identify the exponentially rising and falling terms), and for
\begin{figure}[b]
\vspace{6.1cm}
\includegraphics{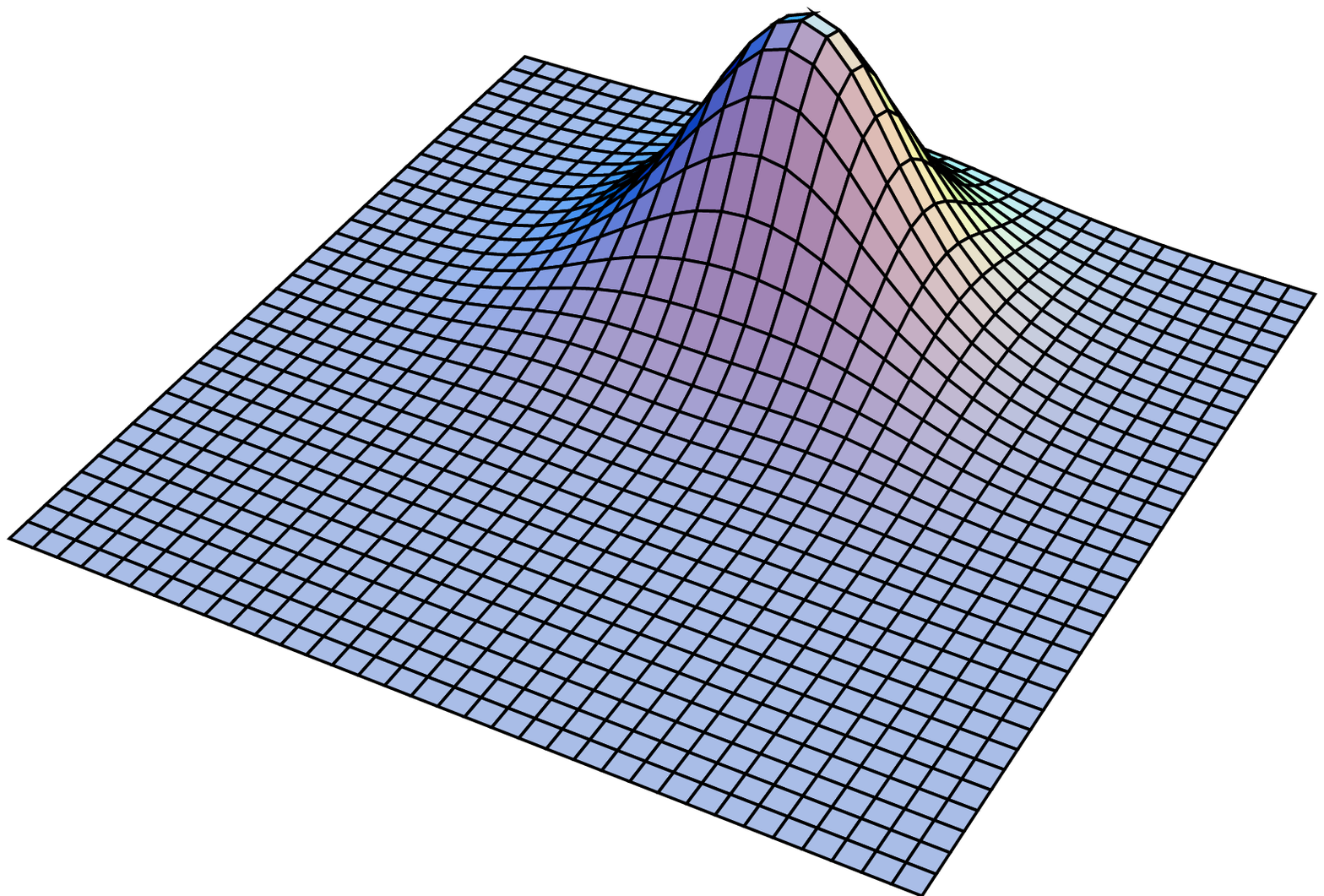}
\includegraphics{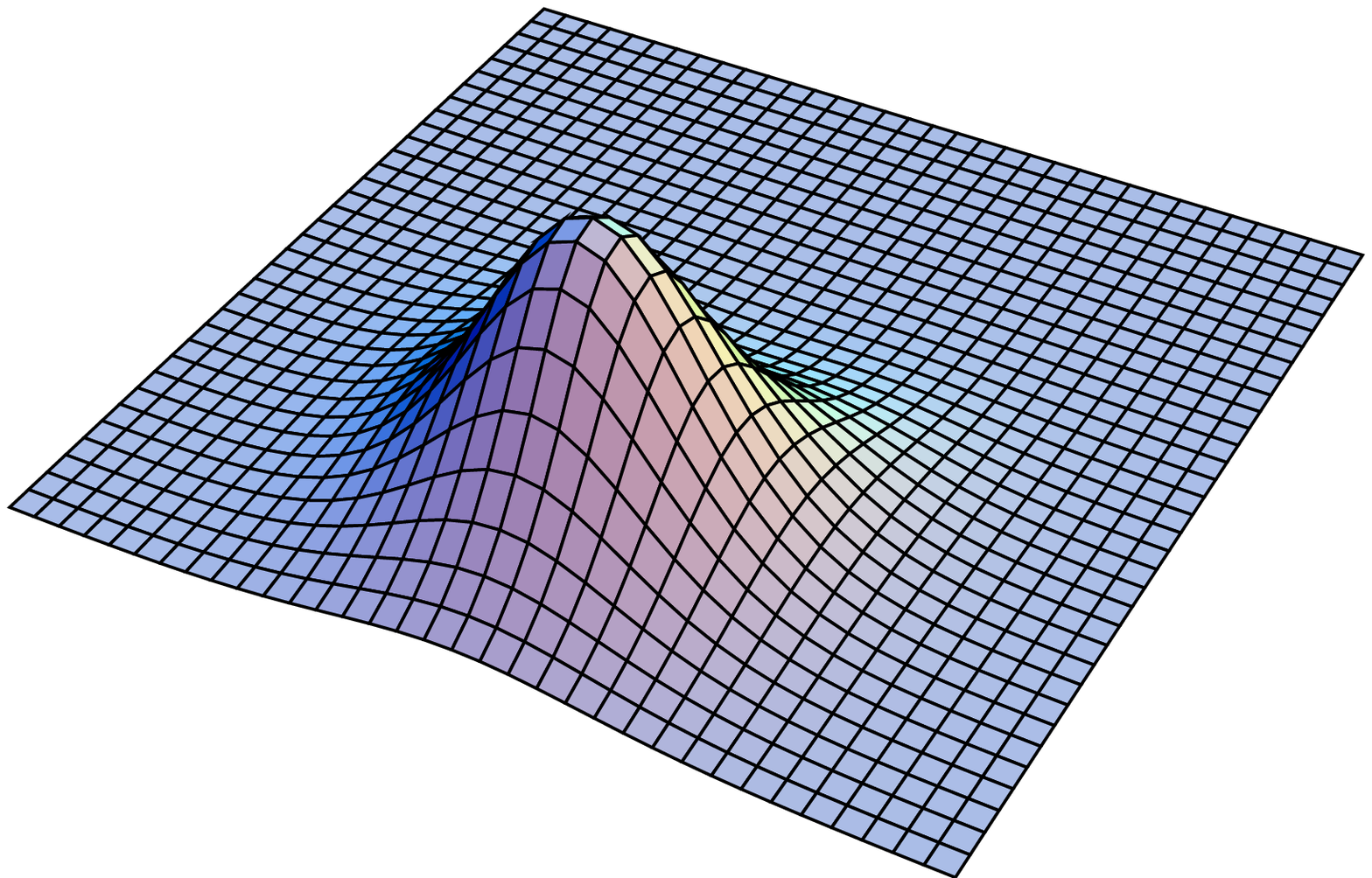}
\includegraphics{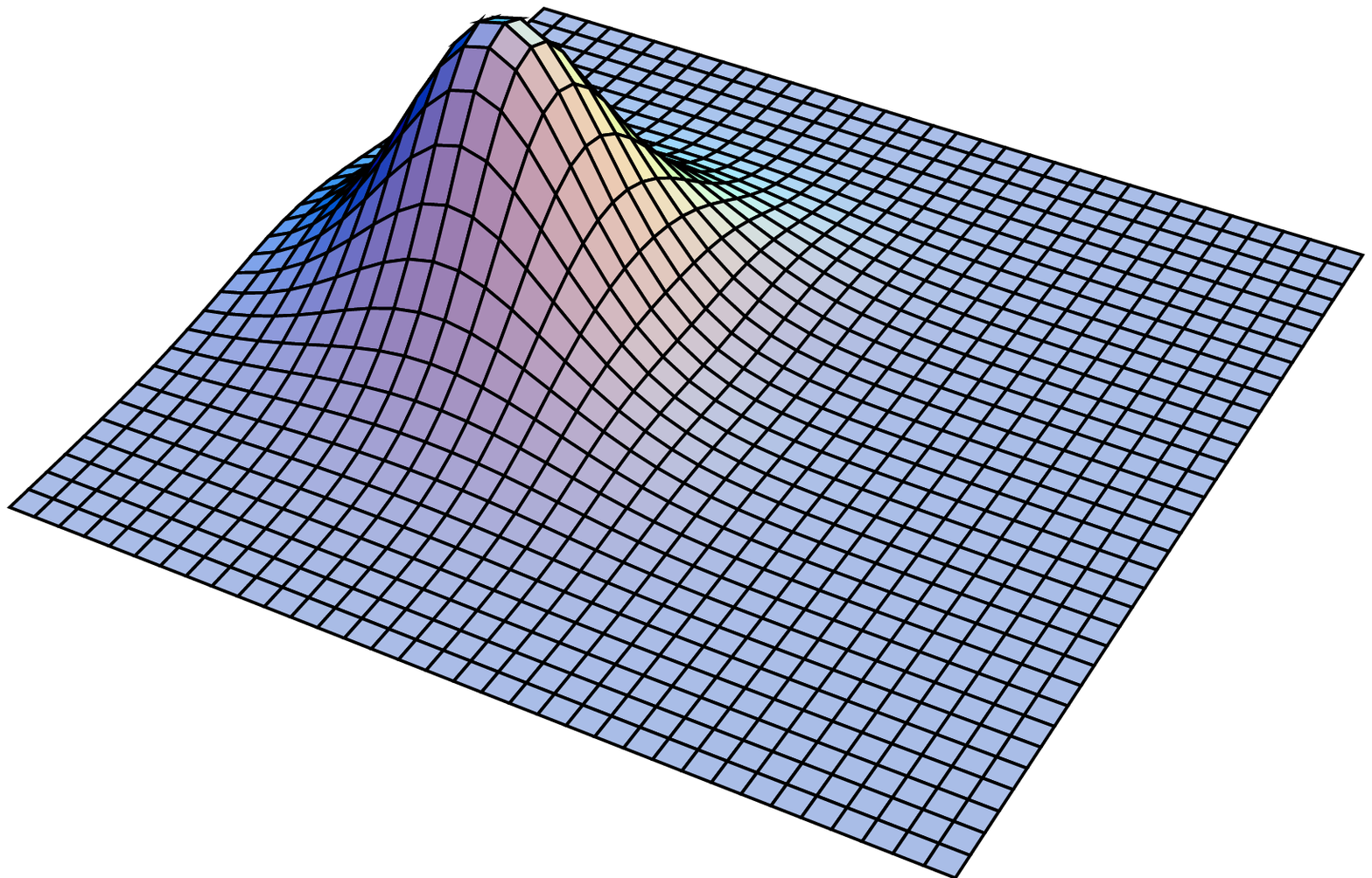}
\includegraphics{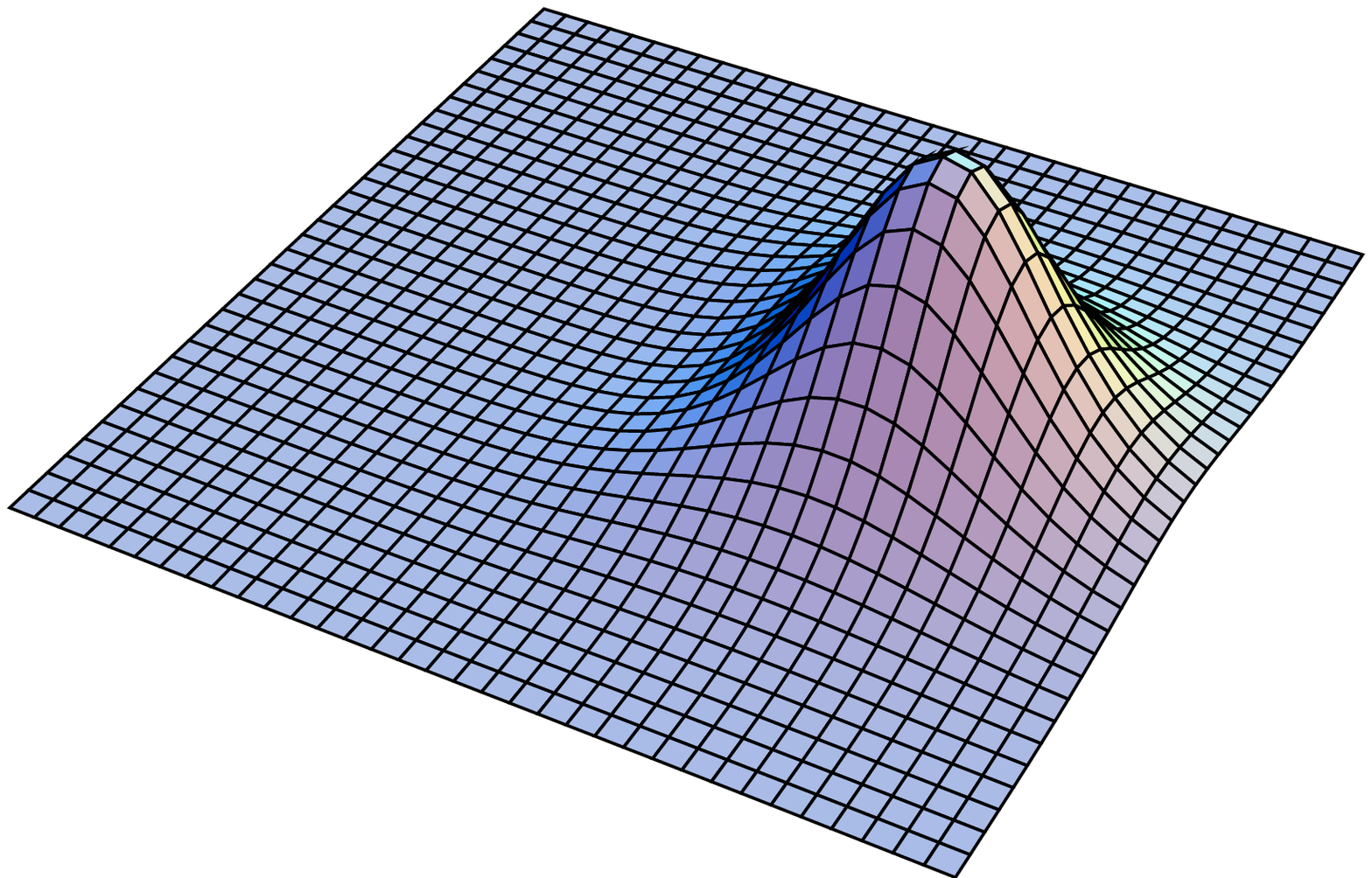}
\includegraphics{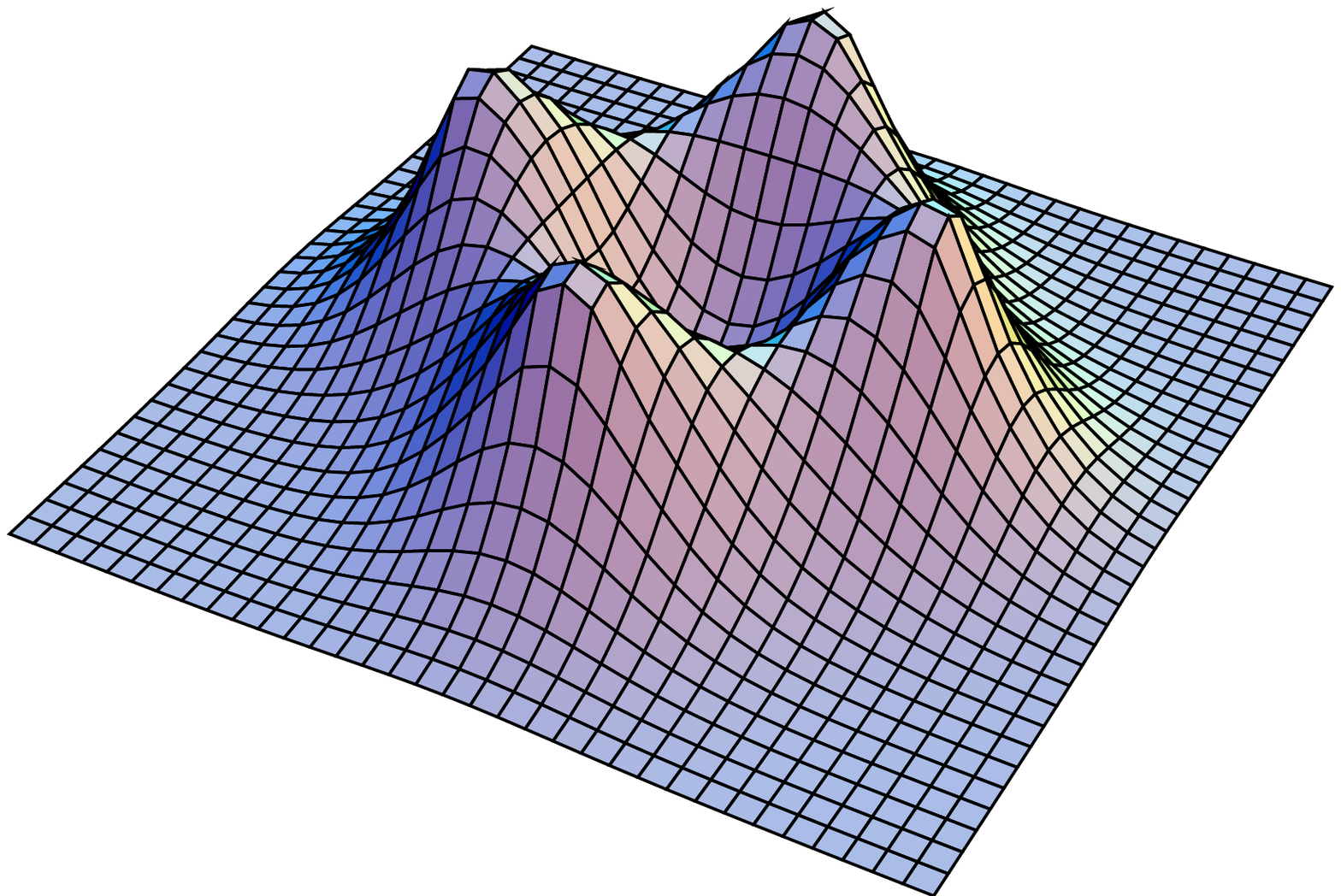}
\caption{In the middle is shown the action density in the plane of the
constituents at $t=0$ for an SU(2) charge 2 caloron with $\tr\,\pl=0$, 
where all constituents strongly overlap. On a scale enhanced by a factor 
$10\pi^2$ are shown the densities for the two zero-modes, using either 
periodic (left) or anti-periodic (right) boundary conditions in the time 
direction.}\label{fig:fig4}
\end{figure}
finding exact solutions through solving the homogeneous Green's 
function~\cite{Us}. We found axially symmetric solutions for arbitrary $k$, 
as well as for $k=2$ two sets of non-trivial solutions for the matching 
conditions that interpolate between overlapping and well-separated 
constituents. For this task we could make use of an existing analytic result 
for charge-2 monopoles~\cite{Hari}, adapting it to the case of carolons. An 
example is shown in Fig.~4.

There has also been some progress on constructing the hyperK\"ahler metric
which approximates the metric for an arbitrary number of calorons. They claim 
that this already gives confinement~\cite{DiPe}. For the implications of 
caloron constituents near the phase transition see~\cite{Bruc}.

\section*{Acknowledgments}

Again I managed to write proceedings, but I needed somewhat more time,
and I am grateful to Matthias Neubert to allow for that and organizing 
a wonderful conference. Also, there are simply too many names, but I 
thank everybody who worked with me.

\end{document}